\documentstyle[aps,prl,multicol]{revtex}

\begin{document}
\title{Spectral Density Functionals for Electronic Structure Calculations}
\author{S. Y. Savrasov and G. Kotliar}
\address{Center for Materials Theory and Department of Physics and Astronomy, Rutgers%
\\
University, Piscataway, NJ 08854--8019}
\date{May 2001}
\maketitle

\begin{abstract}
We introduce a functional of the local spectral electron density which can
be used to to compute the total energy and the local spectral function of
strongly-correlated materials. We illustrate the applicability of the method
by using as an example the long-standing problem of the electronic structure
of metallic plutonium.
\end{abstract}


\begin{multicols}{2}%
%

Electronic structure approaches based on density functional theory (DFT) in
its local density (LDA) and generalized gradient (GGA) approximations \cite
{DFT} have successfully predicted the ground-state properties of
weakly-correlated materials. Furthermore, the Kohn-Sham spectrum of this
approach has proved to be an excellent starting point for calculations of
physical excitation energies using perturbative approaches such as the GW
approximation \cite{Aryas}.

On the other hand, this approach has failed qualitatively in
strongly-correlated systems such as Mott insulators, cuprates, manganites,
and $f$-electron systems. Several directions have been pursued to treat such
systems beyond LDA: The GW method \cite{Aryas}, self-interaction corrected
schemes \cite{Svane}, the LDA+U method \cite{LDA+U} and more recently
LDA+DMFT \cite{AnisKot,Licht}. The LDA+U approach combines the Dynamical
Mean Field Theory (DMFT) \cite{DMFTreview} which has been very successful in
the context of model Hamiltonians, with state--of--the--art band structure
methods. This combination allows a coherent description of the electron in
real and momentum space and leads to the appearance of both Hubbard bands
and quasiparticle bands. Hence, it is well suited to describe
strongly-correlated materials. For recent applications of this method, see 
\cite{Licht,NekrVollhard,PuNature}.

In this Letter we introduce a new functional formulation of the electronic
structure problem which leads to the LDA+DMFT equations and we present an
application of the full implementation of this method to plutonium. In our
generalization of DMFT to the electronic structure problem, our basic idea
is to introduce another relevant variable in addition to the density $\rho $%
, namely the local Green function. The latter is defined by projecting the
full Green function onto a separate subset of correlated ``heavy'' orbitals
distinguished by the index $a$ from a complete set (indexed by $\alpha $) of
orbitals $\chi _{\alpha }({\bf r}-{\bf R})\equiv \chi _{\alpha R}$ of a
tight-binding representation which we assume for simplicity to be
orthogonal. The local Green function is therefore given by a matrix $\hat{G}$
with elements \cite{chitra} 
\begin{eqnarray}
&&{\ }G_{ab}(i\omega ,R)=-\left\langle c_{aR}(i\omega )c_{bR}^{+}(i\omega
)\right\rangle =  \nonumber \\
- &&\int \chi _{a}^{\ast }({\bf r-R})\left\langle \psi ({\bf r},i\omega
)\psi ^{+}({\bf r}^{\prime },i\omega )\right\rangle \chi _{b}({\bf r}%
^{\prime }-{\bf R})d{\bf r}d{\bf r}^{\prime }.  \label{Gab}
\end{eqnarray}
We then construct a functional $\Gamma \lbrack \,\rho ,\hat{G}\,]$ which
gives the exact free energy at a stationary point. This generalizes the
effective action construction used in connection with either pure DMFT
functionals \cite{chitra} or pure density functionals \cite{fukuda}. It
amounts to first considering the partition function of the interacting
electron gas in the presence of a static source coupled to the density and a
dynamic source coupled to the local spectral function, and second, carrying
out a Legendre transformation with respect to those sources. This functional
can be constructed formally in perturbation theory, however its explicit
form is not available just as in density functional theory. The success of
DMFT in model hamiltonians suggests a useful approximation to the exact
functional, namely the LDA+DMFT functional, which allows us to compute both
total energies and spectra.

To describe the new method, it is useful to introduce the notion of the
Kohn-Sham potential $V_{KS}$ and its dynamical analog $\Sigma$. They are
defined as the functions that one needs to add to the kinetic energy matrix
so as to obtain a given density and spectral function of the heavy orbitals
namely:

\begin{equation}
\rho ({\bf r})=T\sum_{\omega }e^{i\omega 0^{+}}\langle {\bf r}|[i\omega
+\nabla ^{2}/2-V_{KS}-\Sigma ]^{-1}|{\bf r\rangle .}  \label{rho1}
\end{equation}
$V_{KS}$ is a function of ${\bf r}$, the chemical potential $\mu $ is set to
zero throghout the paper, and $\Sigma $ is given by 
\begin{equation}
\Sigma \equiv \Sigma ({\bf r},{\bf r}^{\prime },i\omega )=\sum_{abR}\chi
_{a}^{\ast }({\bf r-R})\Sigma _{ab}(i\omega )\chi _{b}({\bf r}^{\prime }{\bf %
-R}),  \nonumber
\end{equation}
Recall that we are truncating to the ``heavy'' subset $\{a\}$ of orbitals.
Then expression (\ref{rho1}) becomes 
\begin{equation}
\rho ({\bf r})=T\sum_{ab}\sum_{\omega {\bf k}}\chi _{a{\bf k}}^{\ast }({\bf r%
})\left[ i\omega -\hat{H}^{{\bf k}}-\hat{\Sigma}(i\omega )\right]
_{ab}^{-1}\chi _{b{\bf k}}({\bf r}),  \label{rhoagain}
\end{equation}
where $H_{ab}^{{\bf k}}=\langle \chi _{a{\bf k}}|-\nabla ^{2}+V_{KS}|\chi _{b%
{\bf k}}\rangle $ is the one-electron hamiltonian in ${\bf k}$-space.

The physical meaning of the dynamical potential $\Sigma $ is parallel to the
meaning of the original Kohn-Sham potential $V_{KS}$: it is the function
that one needs to add to the correlated block of the one-electron
hamiltonian in order to obtain the desired local Green function: 
\begin{equation}
G_{ab}(i\omega )=\sum_{{\bf k}}\,[i\omega -\hat{H}^{{\bf k}}-\hat{\Sigma}%
(i\omega )]_{ab}^{-1}.  \label{Gab2}
\end{equation}
It is the frequency dependence of the dynamical potential which allows us to
treat Hubbard bands and quasiparticle bands on the same footing.

If the exact self energy of the problem is ${\bf k}$-independent, then it
coincides with $\Sigma$. The parallel observation within DFT is: if the self
energy of a model is ${\bf k}$- {\em and} frequency-independent, then the
self energy coincides with the Kohn-Sham potential.

In terms of these quantities, the external potential and the matrix of local
interactions $\hat{U}$, we write down the DMFT+LDA functional: 
\begin{eqnarray}
&&\Gamma _{LDA+DMFT}(\rho ,V_{KS,}\hat{G},\hat{\Sigma})=  \nonumber \\
&&-T\sum_{\omega }e^{i\omega 0^{+}}{\rm Tr}\log [i\omega +\nabla
^{2}-V_{KS}-\Sigma ]  \nonumber \\
&&-\int V_{KS}({\bf r})\rho ({\bf r})d{\bf r}-\sum_{\omega }e^{i\omega 0^{+}}%
{\rm Tr}[\hat{\Sigma}(i\omega )\hat{G}(i\omega )]  \nonumber \\
&&+\int V_{ext}({\bf r})\rho ({\bf r})d{\bf r}+\frac{1}{2}\int \frac{\rho (%
{\bf r})\rho ({\bf r}^{\prime })}{|{\bf r}-{\bf r}^{\prime }|}d{\bf r}d{\bf r%
}^{\prime }+E_{xc}^{LDA}[\rho ]  \nonumber \\
&&+\sum_{R}[\Phi \lbrack \hat{G}]-\Phi _{DC}].  \label{functional}
\end{eqnarray}
$\Phi \lbrack \hat{G}]$ is the sum of the two--particle irreducible local
diagrams constructed with the local interaction matrix $\hat{U}$, and the
local heavy propagator $\hat{G}$. $\Phi _{DC}$ is the so--called double
counting term which subtracts the average energy of the heavy level already
described by LDA. We write it here in its simple form, when only one Slater
integral is included, \i .e. $\Phi _{DC}=\bar{U}\bar{n}(\bar{n}-1)/2$ with $%
\bar{n}=T\sum_{\omega ,ab}G_{ab}(i\omega )e^{i\omega 0^{+}}$ (see Ref. \cite
{LDA+U} for its general form). Expression \ref{functional}ensures that the
Greens function  obtained from its extremization will satisfy the Luttinger 
theorem.  

An explicit form of $\Phi \lbrack \hat{G}]$ is unavailable.
We can calculate it by relating it to the free energy of an atom in
a medium, $W_{at},$ which we describe by an action $S_{at}.$ To do this, we
introduce the second central concept of DMFT: the Weiss field $G_{0}^{-1}$
which represents the quadratic part of the action $\ S_{at}$. 
\begin{eqnarray}
S_{at}[G_{0}^{-1}] &=&\int_{\tau \tau ^{\prime }}\sum_{ab}c_{a}^{+}(\tau
)[G_{0}^{-1}(\tau ,\tau ^{\prime })]_{ab}c_{b}(\tau ^{\prime })+  \nonumber
\\
&&\int_{\tau }\sum_{abcd}U_{abcd}c_{a}^{+}(\tau )c_{b}^{+}(\tau )c_{c}(\tau
)c_{d}(\tau ).  \label{Sat}
\end{eqnarray}
Here, $\hat{G}_{0}^{-1}=\hat{G}_{at}^{-1}-\hat{\Delta}$, where $\hat{G}%
_{at}^{-1}$ describes the quadratic part of the action of the isolated atom,
and $\hat{\Delta}$ describes the bath surrounding the atom. The bath added
to the atomic action produces the desired local Green function i.e. $\langle
c_{a}c_{b}^{+}\rangle _{S_{at}}=G_{ab}.$ We can now relate the sum of local
graphs $\Phi \lbrack \hat{G}]$ to the free energy of the atom in the medium $%
W_{at}=-\log \int \exp [-S_{at}]$ via

\begin{equation}
\Phi \lbrack \hat{G}]=W_{at}[\hat{G}_{0}^{-1}]-{\rm Tr}(\hat{G}_{0}^{-1}-%
\hat{G}^{-1})\hat{G}-{\rm Tr}\log \hat{G}.  \label{FG}
\end{equation}

The functional (\ref{functional}) can be viewed as a functional of four
independent variables, since the stationary condition in the conjugate
fields reproduces the definition of the dynamical potential and the Weiss
field. Extremizing it leads us to compute the Green function $G_{ab}(i\omega
)$ Eq.\ (\ref{Gab2}) with the Kohn-Sham potential entering $H_{ab}^{{\bf k}}$
and with 
\begin{equation}
\Sigma _{ab}(i\omega )=\frac{\delta \Phi }{\delta G_{ab}(i\omega )}-\frac {%
\delta \Phi _{DC}}{\delta G_{ab}(i\omega )},  \label{sigmaabw}
\end{equation}
which identifies $\Sigma (i\omega )$ as the self energy of a generalized
Anderson impurity model \cite{DMFTreview} in a bath. The ``impurity" is
characterized by a matrix of (local) levels 
\begin{equation}
\epsilon _{ab}=\sum_{{\bf k}}H_{ab}^{{\bf k}}  \label{eab0}
\end{equation}
and is hybridized with the medium via the hybridization function $\Delta
_{ab}(i\omega)$. The matrix $\delta \Phi _{DC}/\delta G_{ab}(i\omega)= 
\newline
\bar{U}(\bar{n}-1/2)$ takes into account the double counting effects of the
Coulomb interaction already contained in matrix $H_{ab}^{{\bf k}}.$ The
hybridization function $\Delta _{ab}(i\omega )$ obeys the self-consistency
condition

\begin{eqnarray}
&&i\omega -\epsilon _{ab}-\Delta _{ab}(i\omega )-\Sigma _{ab}(i\omega )= 
\nonumber \\
&&\left[ \sum_{{\bf k}}[i\omega -\hat{H}^{{\bf k}}-\hat{\Sigma}(i\omega
)]^{-1}\right] _{ab}^{-1}  \label{delta}
\end{eqnarray}
The self-consistent DMFT loop with respect to $\hat{\Delta}(i\omega )$ and $%
\hat{\Sigma}(i\omega )$ amounts to: (i) choosing some input hybridization $%
\hat{\Delta}(i\omega )$, (ii) solving the Anderson impurity model with that
hybridization and extracting the impurity Green function and self energy $%
\hat{\Sigma}(i\omega )$, (iii) computing the on-site crystal Green function
according to Eq.\ (\ref{Gab2}), and (iv) recovering a new hybridization
function $\hat{\Delta}(i\omega )$ from Eq.\ (\ref{delta}).

Since the charge density depends on $\hat{\Sigma}(i\omega)$, another
self-consistent loop is required to obtain the extremum of Eq.\ (\ref
{functional}). This second loop updates the charge density once the DMFT
loop delivers the self energy; this modifies the hopping integrals in the
one-electron hamiltonian $\hat{H}^{{\bf k}}$. The new hoppings set up a new
set of impurity levels $\epsilon_{ab}$. We see that the solution of eqs.(\ref
{sigmaabw})--(\ref{delta}) should be carried out in a double iterational
loop: the DMFT loop finds $\hat{\Sigma}(i\omega )$ for given $\rho $ while
the global one (DFT like) updates $\rho.$ Within the DMFT technology, the
implementation of this global loop is an essential new development of the
present work which allows the determinatoion of a central quantity of the
solid -- its total energy.

We now illustrate the method applied to the electronic structure of Pu \cite
{PuBook}. Previous LDA\cite{Pucalc1,Pucalc2} or LDA+U \cite{PuLDA+U} or
model \cite{Eriksson} calculations have not been able to describe the
existence of both a compressed $\alpha $ phase and a low-density $\delta $
phase, nor to provide a qualitative description of its photoemission spectra 
\cite{PuExp1}. We have recently reported the energy vs volume curve and the
one-electron spectrum for a value of $U$ which was estimated to correspond
to that for $\delta $ Pu \cite{PuNature}. Here, we study in detail the
energy as a function of the parameter $U$ and give a comparative analysis of
the spectra in both $\alpha$ and $\delta $ phases We also report
calculations on a bcc structure corresponding to the higher temperature $%
\epsilon$ phase of Pu.

We have implemented the spectral density functional approach described above
using the linear muffin tin orbital (LMTO) method for electronic structure
calculations. We used a generalized non-orthogonal tight-binding LMTO\
representation \cite{OKA} which is most suited for describing the set of
localized orbitals. The non-orthogonality corrections enter the definition
for the Green function (\ref{Gab2}) in the form of an overlap matrix. To
solve the impurity model for multi-orbital case of Pu we use a method which
interpolates the self energy calculated from the Gutzwiller approximation 
\cite{Gutz} at low frequencies and a Hubbard-1-like approximation \cite{Hub}
at high frequencies \cite{GutzHub}. Spin--orbit coupling effects are
generally important for actinide compounds and have been included in the
calculation for Pu. The ``full potential'' terms have been neglected in the
calculations through the utilization of the atomic sphere approximation with
a one-kappa LMTO basis set \cite{FPLMTO}.

The necessary $k$-space integrals for evaluating Green functions and charge
densities have been carried out using the tetrahedron method with on a
reciprocal grid (8,8,8). The $\delta$ and $\epsilon$ phases of Pu have
simple fcc and bcc structures respectively, while the $\alpha $ phase has a
complicated lattice with 16 atoms per unit cell. We determined the ground
state energy as a function of the atomic volume using fcc and bcc
geometries. Finally, the approach requires the effective Coulomb interaction
between $f$ electrons described by the Hubbard parameter $U$. Various
estimates exist in the literature on the values of $U$ in plutonium which
indicate that the average interaction among $f$ electrons is around 4 eV 
\cite{PuU}.

We now report our calculated total energies for Pu in the fcc structure. The
temperature is fixed at $600K$, where the $\delta $ phase is stable. To
illustrate the importance of correlations we perform the calculations for
several values of $U$ varying it from 0 and 4 eV. The results are shown in
Fig.\ 1: The $U$=0 (GGA) curve shows a minimum at $V/V_{\delta }$=0.72,
close to the volume of the low-temperature $\alpha $ phase. Here $V_{\delta
} $ is the observed $\delta $--phase volume. We expect that correlations
should be less important for the compressed lattice in general, but there is
no sign whatsoever of the $\delta $ phase in the $U$=0 calculation. Dramatic
changes occur when $U$ is about 4 eV. The curves show the possibility for a
double minimum while the details depend sensitively on the actual value of $%
U $. We find that for $U=$ 3.8 eV, the minimum occurs near V/V$_{\delta }$%
=0.80 which corresponds to the volume of the $\alpha $ phase. When $U$
increases by 0.2 eV the minimum occurs at V/V$_{\delta }$=1.05 which
corresponds to the volume of the $\delta $ phase, in close agreement with
experiment. Clearly, since the energies are so similar, we may expect that
as temperature decreases, the lattice undergoes a phase transition from the $%
\delta $ phase to the $\alpha $ phase with a remarkable decrease of the
volume by 25\%.

We repeated our calculations for the bcc structure using the temperature $%
T=900K$ where the $\epsilon $ phase is stable. Fig.\ 1 shows these results
for $U=$ 4 eV with a location of the minimum at around V/V$_{\delta }$=1.03.
While the theory has a residual inaccuracy in determining the $\delta $ and $%
\epsilon $ phase volumes by a few percent, a hint of volume decrease with
the $\delta \rightarrow \epsilon $ transition is clearly reproduced. Thus,
our first-principles calculations reproduce the main features of the
experimental phase diagram of Pu.

We now report our calculated spectral density of states for the fcc
structure using the volumes V/V$_{\delta }$=0.8 and V/V$_{\delta }$=1.05
corresponding to the $\alpha $ and $\delta $ phases. Fig.\ 2 shows the
results of our dynamical mean-field calculations as compared to the
LDA--type calculations. We predict the appearance of a strong quasiparticle
peak near the Fermi level which exists both in the $\alpha $ and $\delta $
phases. Recent advances have allowed the experimental determination of these
spectra, and our calculations are in accord with these measurements \cite
{PuExp1}. The width of the quasiparticle peak in the $\alpha $ phase is
found to be larger by 30 per cent compared to the width in the $\delta $
phase. This indicates that the low-temperature phase is more metallic, i.e.
it has larger spectral weight in the quasiparticle peak and smaller weight
in the Hubbard bands.

In conclusion, we have developed a new method for electronic structure
calculations which allows the simultaneous evaluation of the total energy
and the local electronic spectral density. We applied the method to study
the phase diagram of Pu and have reproduced several of its salient features.
While we have not yet included the electronic and phonon entropy terms which
are difficult to evaluate numerically, many properties of realistic strongly
correlated systems which are related to the total energy can now be explored
using LDA+DMFT.

The authors are indebted to E. Abrahams, A. Arko, J. Joyce and J. Thompson
for helpful discussions. This research   was supported by the US\ DOE,
grant No.
DE-FG02-99ER45761,   by Los Alamos National Laboratory under subcontract
No. 25124-001-012K, and by the NSF DMR-0096462.

\centerline{\bf FIGURE CAPTIONS}

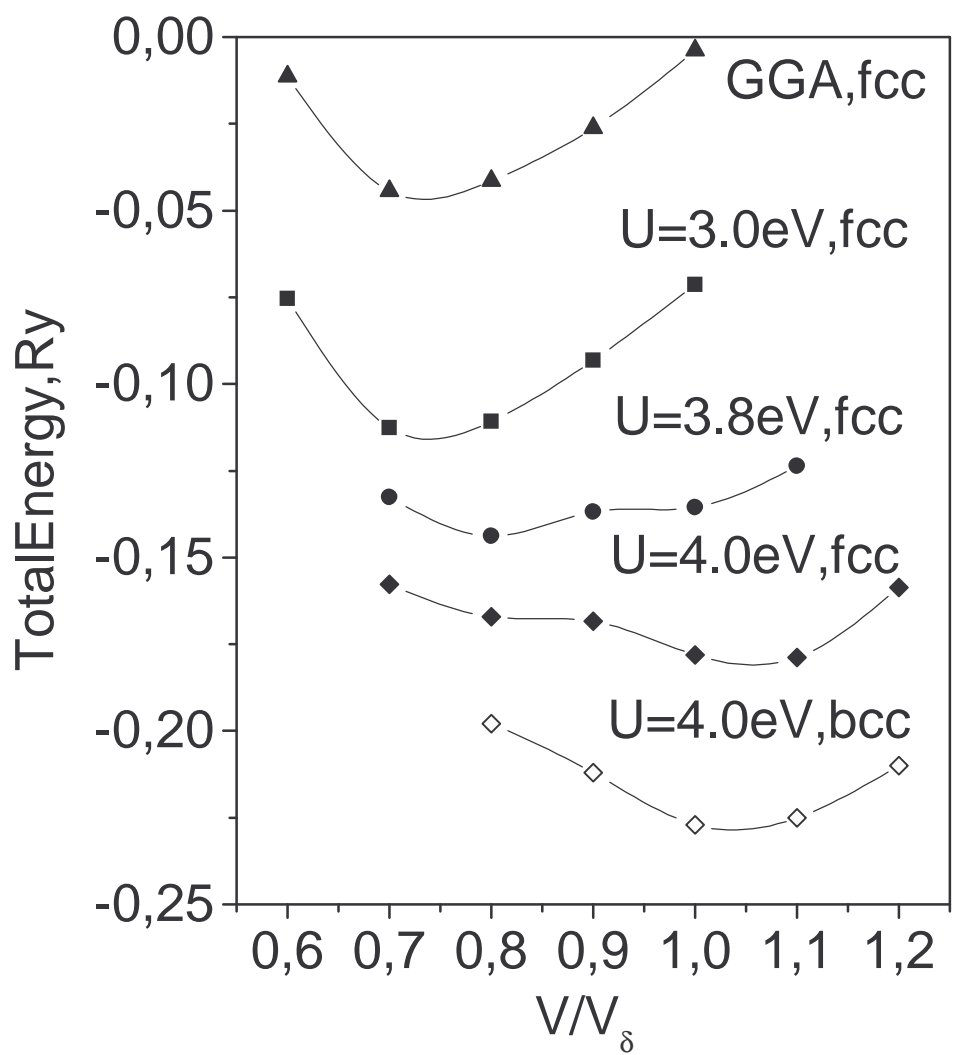
Fig. 1. Total energy as a function of volume in Pu for different values of $%
U $ calculated using the LDA+DMFT approach. Data for the fcc lattice are
computed at T=600K, while data for the bcc lattice are given for T=900K.

Fig. 2. Comparison between photoemission spectra of $\delta -$Pu \cite
{PuExp1} (circles) and calculated densities of states using the LDA+DMFT
approach: the data for $V/V_{\delta }=1.05,\,U=4.0$ eV (full line) and the
data for $V/V_{\delta }=0.80,\,\,U=3.8$ eV (dashed line) correspond to the
volumes of the $\delta $ and $\alpha $ phases respectively. The result of
the GGA\ calculation (dotted line) at $V/V_{\delta }=1\,\,(U=0)$ is also
given.

\end{multicols}%
%

\end{document}